\documentstyle{article}
\begin{document}

\begin{titlepage}

\begin{flushleft}
TU-529
\end{flushleft}

\begin{center}
{\LARGE \bf
 Black Holes \\
\ \\
and\\
\ \\
Two-Dimensional Dilaton Gravity
}
\end{center}
\ \\
\ \\
\begin{center}
\Large{
T.Futamase${\ }^{\flat}$, M.Hotta${\ }^{\natural}$ 
and Y.Itoh ${\ }^{\flat}$ 
 }\\
{\it 
${\ }^{\natural}$ Department of Physics, Faculty of Science, Tohoku University,\\
Sendai 980-77, Japan\\
${\ }^{\flat}$ Astronomical Institute, Faculty of Science, Tohoku University,
Sendai 980-77, Japan
}
\end{center}

\begin{abstract}

We study the conditions for 2-dimensional dilaton gravity models to 
have dynamical formation of black holes and construct all such models.  
Furthermore we present a parametric representation of 
the general solutions of the black holes. 

\end{abstract}

\end{titlepage}

\section{Introduction}

\ \\

The two-dimensional dilaton gravity has attracted 
much attention in the context not only of string theory but also 
of the Hawking radiation.

It turns out that even in two dimensions the event horizon can be 
dynamically formed and the thermal radiation is emitted from the hole 
if the quantum effect is taken into account\cite{CGHS,DG}.
It is widely believed that the physical essence of the two-dimensional 
Hawking radiation is equivalent to the four-dimensional physics.

The dilaton gravity in two dimensions does not have a unique action 
and there exist an infinite number of variations of the theories. 
Thus it is quite crucial to extract aspects 
independent of the detail of the models.

Much effort has been already made to solve the most general 
dilaton gravity model and many progresses have been achieved
in this direction\cite{BO,F,LPOS,M,LK}. 
Especially the general static solutions are formally obtained 
using the conformal transformation technique\cite{LK}. 
 
However it is not fully addressed the question about what kinds of 
the dilaton gravity models survive after imposing 
the following requirements which is quite reasonable to 
expect to hold. 
Firstly the model should have the flat vacuum solution 
in order to describe the state before the black-hole formation. 
Secondly the model should admit a family of static black-hole solutions 
with regular horizons. The mass of the hole can get heavier without bound. Finally the black holes should be created in process of the gravitational 
collapse. For example the infalling matter shock-wave will bring about 
the formation of a black hole.

In this paper we address and solve this question. Namely we construct 
all the dilaton gravity theories containing at most second derivatives of the fields in the action which satisfy the above requirements.
Moreover we give a parametric representation of the general 
black-hole solutions.

This paper is organized as follows. 
We give a short review on the most general model of dilaton gravity 
in section 2. 
In section 3 we analyze the conditions for the existence of the flat vacuum 
as well as a black hole, and then construct all the theories which admit at 
least one black-hole solution by imposing appropriate boundary condition at the 
black hole horizon.
It is shown that the theories can be completely labeled using 
two parameters $g$ and $\mu$.
In section 4 we discuss the dynamical formation of the black hole.
In this argument the parameter $\mu$ is identified as the black-hole mass.
In section 5 we obtain general black-hole solutions with larger masses than $\mu$. 
Furthermore it will be shown that the parameter $g$ is fixed  
if the event horizon does not vanish for the 
arbitrarily large mass of the hole.  
Consequently it will be argued that all the theories 
satisfying the natural criteria are completely fixed  by 
giving an arbitrary input black-hole configuration and its mass  $\mu$.
 In the section 6 and 7 we discuss two examples in some detail.

\section{Two-Dimensional Dilaton Gravity}

\ \\

Let us first write down the action of the dilaton gravity 
including only second derivative terms and 
the dilatonic cosmological term.
\begin{eqnarray}
S&=&\frac{1}{2\pi}\int d^2 x \sqrt{-g}
     \left( F(\phi) R + 4G(\phi)(\nabla \phi)^2 + 4\lambda^2 U(\phi)
            -\frac{1}{2}(\nabla f)^2 \right).\label{1}
\end{eqnarray}
Here $g_{ab}$ is the two-dimensional metric and $\phi$ is 
the dilaton field.
The forms of the functions $F$, $G$ and $U$ are arbitrary so far, 
but will be specified later to admit solutions 
describing black formation process.
The constant $\lambda$ has one mass dimension.
A massless matter field $f$ is included in the above action since 
we shall consider the process of gravitational collapse in later section.

Now we adopt the conformal gauge fixing for the metric.
$$
x^{\pm} = t \pm r .
$$
\begin{eqnarray}
ds^2 = -e^{2\rho} dx^+ dx^-
\label{2}
\end{eqnarray}
Note that there remains a residual gauge freedom in equation
 (\ref{2}), that is, the conformal gauge transformation:
\begin{eqnarray}
&&
{x'}^\pm ={x'}^\pm (x^\pm ),
\\
&&
\rho' =\rho -\frac{1}{2}\ln
\left[\frac{d{x'}^+}{dx^+}\frac{d{x'}^-}{dx^-}\right].
\end{eqnarray}
We do not fix this gauge freedom and change the coordinates 
at our convenience.

The connections and the scalar curvature can be simply calculated 
From the above metric.
\begin{eqnarray}
\Gamma^+_{++} &=& 2\partial_+ \rho ,
\nonumber\\
\Gamma^-_{--} &=& 2\partial_- \rho ,
\nonumber
\end{eqnarray}
$$
R= 8 e^{-2\rho} \partial_+ \partial_- \rho .
$$

The equations of motion are derived by taking variation of 
the action (\ref{1}). In the conformal gauge they are written as
\begin{eqnarray}
&&
2F' \partial_+ \partial_- \rho
+8G\partial_+ \partial_- \phi
+4G' \partial_+ \phi \partial_- \phi
+\lambda^2 U' e^{2\rho}  =0,\label{3}\\
&&
\partial_+ \partial_- F  
+\lambda^2 U e^{2\rho} =0,\label{4}\\
&&
\partial_{\pm}^2 F -2\partial_{\pm} \rho\partial_{\pm}F
-4G(\partial_{\pm} \phi )^2
= -\frac{1}{2}(\partial_\pm f)^2.\label{5}
\end{eqnarray}
The equation (\ref{5}) is the first constraint of the system and 
the equation (\ref{3}) is automatically satisfied 
if equations (\ref{4}) and (\ref{5}) hold.

For static solutions without matter
the equations (\ref{3}) $\sim$ (\ref{5}) can be rewritten 
in the following forms.
\begin{eqnarray}
&&
F' \ddot{\rho} +4G\ddot{\phi} =
2\lambda^2 U' e^{2\rho} -2G'\dot{\phi}^2,
\label{6}\\
&&
 F'\ddot{\phi}=4\lambda^2 U e^{2\rho}
-F'' \dot{\phi}^2,
\label{7}\\
&&
\lambda^2 U e^{2\rho}
= G\dot{\phi}^2
+
\frac{1}{2}F'\dot{\rho}\dot{\phi}
\label{8}
\end{eqnarray}
Here the prime means the derivative with respect to $\phi$
and the dot to the spacelike variable $r$.

\section{Existence of the Flat Vacuum State and a Black Hole Solution}

\ \\

In order to restrict general two-dimensional dilaton gravity 
described in the previous section, we first 
consider the condition for the existence of the flat vacuum solution:
\begin{eqnarray}
&&
\rho=C=const,
\label{9}\\
&&
\phi=\phi_{vac} (r).
\label{10}
\end{eqnarray}

Substituting (\ref{9}) and (\ref{10}) into the equations of motions
 (\ref{6}) $\sim$ (\ref{8}), we get the necessary 
condition of the vacuum existence as follows.
\begin{eqnarray}
\frac{U'}{U}=\frac{G'}{G} -2\frac{F''}{F'} +8\frac{G}{F'} .
\label{11}
\end{eqnarray}
Furthermore it is easily seen that the condition (\ref{11}) 
is used to show that the theory has the flat vacuum solution and  
the dilaton field in the vacuum state is then obtained by integrating
the following differential equation:
\begin{eqnarray}
\frac{d\phi }{dr} =\lambda e^C 
\left(
\frac{U(\phi)}{G(\phi )}
\right)^{1/2} .
\label{12}
\end{eqnarray}

Next we consider the  condition for the existence of a black hole 
solution. 
The black-hole configuration can be generally expressed, using
the conformal gauge transformation, as 
\begin{eqnarray}
&&
\phi =\phi(r)
\label{13}\\
&&
\rho(r\sim-\infty)
=\lambda r +\rho_o +\rho_1 e^{2\lambda r} +\rho_2 e^{4\lambda r}
+\cdots
\label{14}\\
&&
\lim_{r\rightarrow\infty} \rho =\rho(\infty) =finite .
\label{15}
\end{eqnarray} 
Otherwise the function $\rho(r)$ is chosen arbitrary except that 
$\rho(\infty) > \rho(r)$ which is needed to ensure the regularity of the 
dilaton field.   
In this expression the event horizon stays at $r=-\infty$ 
and the scalar curvature on the horizon  is $-8\lambda^2 \rho_1$.
It is of course possible to investigate inside the horizon
 by changing the coordinates like
$$
x^\pm = \pm\frac{1}{\lambda}e^{\pm\lambda X^\pm}.
$$
Then the horizon is mapped into null surfaces $X^+ =0$ and $X^- =0$.

Let us assume  the configuration  to be a solution of the equations of motion
(\ref{6}) $\sim$ (\ref{8}).
If we introduce three functions of $r$:
\begin{eqnarray}
&&
F=F(\phi (r) ),
\\
&&
\bar{G} =\dot{\phi }^2 G(\phi (r)),
\\
&&
U=U(\phi (r)),
\end{eqnarray}
then the equations of motion are rewritten as
\begin{eqnarray}
&&
\dot{F} \ddot{\rho}  =
2\lambda^2 \dot{U} e^{2\rho } -2\dot{\bar{G}},
\label{16}\\
&&
\ddot{F}=4\lambda^2 U e^{2\rho },
\label{17}\\
&&
\lambda^2 U e^{2\rho }
=\bar{G}+\frac{1}{2}\dot{F}\dot{\rho}.
\label{18}
\end{eqnarray}

It is quite essential to realize the fact that giving a black-hole solution
imposes restrictions on the theory. The form
of $\rho(r)$ which satisfies the boundary condition (\ref{14}) $\sim$ (\ref{15}) leads us to explicit forms of $F$, $G$ and $U$.

Let us prove this statement.
From the equations of motion (\ref{17}) and (\ref{18})
we can express $\bar{G} (r)$ and $U(\phi(r))$ by $F(\phi (r))$
and $\rho(r)$.
\begin{eqnarray}
&&
\bar{G}
=\frac{1}{4}
\left[
\ddot{F} -2\dot{\rho} \dot{F} 
\right]
\label{20}\\
&&
U
=\frac{e^{-2\rho} }{4\lambda^2}
\ddot{F}
\label{21}
\end{eqnarray}
The functions $F$, $G$ and $U$ are also subject to
the vacuum-existence condition (\ref{10}).  
The equation (\ref{10}) is reexpressed as
\begin{eqnarray}
\frac{\dot{U}}{U }-\frac{\dot{\bar{G}}}{\bar{G}}
+2\frac{\ddot{F}}{\dot{F}}=8\frac{\bar{G}}{\dot{F}}
\label{22}
\end{eqnarray}
By substituting (\ref{20}) and (\ref{21}) into (\ref{22})
we obtain the equation to determine $F(\phi(r))$ from $\rho(r)$. 
\begin{eqnarray}
\frac{F^{(3)}}{\ddot{F}}
-2\frac{\ddot{F}}{\dot{F}}
=
\frac{\ddot{\rho}}{\dot{\rho}}-2\dot{\rho}.
\label{23}
\end{eqnarray}

The differential equation (\ref{23}) can be straightforwardly
integrated and we get the general solution as follows.
\begin{eqnarray}
&&
\dot{F} =\frac{2g \mu}{ e^{-2\rho }-g^2},
\label{24}\\
&&
F=F_o +\int \frac{2g \mu }{ e^{-2\rho }-g^2} dr
\label{25}
\end{eqnarray}
where $\mu$ , $g$ and $F_o$ are integration constants. 
Note that in the case that $g \to 0$ with $g\mu$ fixed the solution 
(\ref{24}) takes the form ${\dot F} = A e^{2\rho}$,  
but it leads to ${\bar G} = 0$ which is not interesting in the present discussion. Thus we will not consider such a case henceforce.  

The integration constant $F_o$ above does not generate any physical effect
because the variation of
$$
\int \sqrt{-g}F_o R d^2 x
$$
in the action yields just a boundary term and does not contribute
 to the equations of motion.

Substituting (\ref{24}) into (\ref{20}) and (\ref{21})
yields the following expressions.
\begin{eqnarray}
&&
\bar{G}=\frac{g\dot{\rho}}{4 \mu }
\left[ \frac{2g \mu}{ e^{-2\rho }-g^2} \right]^2
\label{26}\\
&&
U =\frac{g \mu }{\lambda^2}
\frac{\dot{\rho} e^{-4\rho} }{[ e^{-2\rho} -g^2]^2}
\label{27}
\end{eqnarray}
The equations (\ref{25}),(\ref{26}) and (\ref{27}) are the required 
expressions.

It is notable that $\phi(r)$ is not yet specified 
though we have consumed all the information of the equations 
of motion and the vacuum-existence condition. This is 
reflected by the fact that dilaton-field redefinition:
$$
\tilde{\phi} =\tilde{\phi} (\phi)
$$
does not change the theory itself. 
The functional form of $\phi(r)$ depends on the definition 
adopted by the dilaton field $\phi$ and thus has no physical meaning.
On the other hand the $r$-dependences of $F$, $\bar{G}$ and $U$ 
are independent of the dilaton-redefinition and real physical quantities.

We can take the form of $\phi(r)$ as we wish and 
impose  
\begin{eqnarray}
\dot{\phi} (r)>0.
\label{28}
\end{eqnarray}
Then $r$ is solved by $\phi$ keeping one-to-one correspondence.
\begin{eqnarray}
r=r(\phi).
\end{eqnarray}
Using this and equations (\ref{25})$\sim$(\ref{27}), 
we reconstruct $F(\phi)$, $G(\phi)$ and $U(\phi)$ 
as follows.
\begin{eqnarray}
&&
F =F(\phi(r(\phi)))
\\
&&
G=\frac{\bar{G}(r(\phi))}{\dot{\phi}(r(\phi))^2}
\\
&&
U =U(\phi(r(\phi)))
\end{eqnarray}

As easily understood from the action (\ref{1}), 
the inverse of $F$ can be identified as 
the gravitational constant. Thus in the normal situation
it is natural to take
\begin{eqnarray}
&&
F\geq 0,
\\
&&
\dot{F}>0.
\end{eqnarray}
This property of $F$ enables us to define the dilaton field
 to satisfy 
$$
F(\phi) =\phi,
$$
holding equation (\ref{28}). 
We adopt this definition for $F$ in the following sections.

\section{Gravitational Collapse Solution}

\ \\

The theory we discussed in the section 4 
 has the vacuum solution $(\rho_{vac} (r) =const, \phi_{vac}(r))$
and a black-hole solution $(\rho(r) ,\phi(r))$. 
From these two solutions we can construct a solution which 
describes the dynamical formation of a black hole caused by a
shock-wave type source:
$$
\frac{1}{2}(\partial_+ f )^2 = \varepsilon \delta (x^+ -z^+ ) .
$$

The spacetime is in the vacuum state for $x^+ < z^+ $ and 
a black hole appears for $x^+ > z^+$. Thus the metric of this process may be written as
\begin{eqnarray}
ds^2=-dx^+ dx^-
\left(\Theta(x^+ -z^+ )
 e^{2\rho\left(r=\frac{1}{2}(x^+ - x^- ) \right) }
+\Theta (z^+ -x^+ )
e^{2\rho\left(r=\frac{1}{2} (z^+ -x^- ) \right) }
\right).
\nonumber
\end{eqnarray}
where $\Theta$ is the step function.

For the vacuum portion there exist  the canonical coordinates
$\sigma^\pm$ where the metric and dilaton can be expressed as follows.
\begin{eqnarray}
&&
ds^2 = -e^{2C}d\sigma^+ d\sigma^- = e^{2C}(-dT^2 +dX^2)
\\
&&
\frac{d\phi_{vac} }{dX} =\lambda e^C 
\left(
\frac{U(\phi_{vac} )}{G(\phi_{vac} )}
\right)^{1/2}
\label{29}
\end{eqnarray}

Integration of the equation (\ref{29}) together 
with (\ref{26}) and (\ref{27}) yields the expression for $X$ 
in terms of $r$. 
\begin{eqnarray}
X= g e^{-C} \int^{r(\phi_{vac} )} e^{2\rho (r)} dr.
\end{eqnarray}

It is also easy to write down explicitly 
the coordinate transformation between $x^\pm$ and $\sigma^\pm$.
\begin{eqnarray}
&&
\sigma^+ =x^+ ,
\\
&&
\sigma^- =\int \exp\left[
2\rho\left( r=\frac{1}{2} (z^+ -x^- )\right)-2C \right] dx^- .
\end{eqnarray}

Using the relation:
\begin{eqnarray}
 X=\frac{1}{2} (x^+ -\sigma^- (x^-) ) ,
\end{eqnarray}
we get a non-trivial relation:
\begin{eqnarray}
\frac{1}{2}(x^+ -\sigma^- (x^-))
= g e^{-C} 
\int^{r(\phi_{vac} (x^+ ,x^-)  )} e^{2\rho (r)} dr .
\label{30}
\end{eqnarray}

Differentiation of equation (\ref{30}) with respect to $x^\pm$ 
 yields
\begin{eqnarray}
&&
\frac{\partial\phi_{vac}}{\partial x^+}
=e^C 
\frac{ \mu}{1-g^2 e^{2\tilde{\rho}(x^+ ,x^-)} },
\label{31}\\
&&
\frac{\partial\phi_{vac}}{\partial x^-}
=- e^{-C+2\rho\left(r=\frac{1}{2}(z^+ -x^-) \right)} 
\frac{ \mu }{1-g^2 e^{2\tilde{\rho}(x^+ ,x^-)} },
\label{32}\\
&&
\tilde{\rho}(x^+ ,x^-)
\equiv 
\rho(r=r(\phi_{vac} (x^+ ,x^- )))
\label{33}
\end{eqnarray}
where we used ({\ref{24}) for $F = \phi$.

The shock-wave contribution appears in $++$ part of the equations
 of motion (\ref{5}) as follows.
\begin{eqnarray}
\partial_{+}^2 \phi -2\partial_{+} \rho\partial_{+}\phi
-4G(\partial_{+} \phi )^2
=-\varepsilon \delta (x^+ -z^+ ).
\end{eqnarray}

From this a junction constraint must be imposed.
\begin{eqnarray}
\frac{\partial \phi}{\partial x^+ }|_{x^+ =z^+} 
-\frac{\partial \phi_{vac}}{\partial x^+ }|_{x^+ =z^+}
=-\varepsilon
\label{34}.
\end{eqnarray}

On the other hand the remaining part of the equation (\ref{5}) leads 
to a continuity condition:
\begin{eqnarray}
\frac{\partial \phi}{\partial x^- }|_{x^+ =z^+} 
-\frac{\partial \phi_{vac}}{\partial x^- }|_{x^+ =z^+}
=0.
\label{35}
\end{eqnarray}

Noticing 
$$
\partial_\pm \phi =\pm \frac{1}{2}\frac{d\phi}{dr}
=\pm \frac{g\mu}{e^{-2\rho(r)} -g^2}
$$
and substituting the equations (\ref{31}) and (\ref{32}) into the 
equations (\ref{34}) and (\ref{35}), the following relations are obtained.
\begin{eqnarray}
&&
\varepsilon =\frac{\mu}{g}
\label{36}
\\
&&
e^{2C} =\frac{1}{g^2}
\label{37}\\
&&
\tilde{\rho}(z^+ ,x^-) =\rho\left(r=\frac{1}{2}(z^+ -x^-) \right)
\label{38}
\end{eqnarray}
The last equation (\ref{38}) is automatically satisfied
by the definition (\ref{33}).

From equations (\ref{36}) and (\ref{37}) 
the total mass of the collapsing body is evaluated as
$$
M=e^{-C} \varepsilon =\mu.
$$
Thus the parameter $\mu$ is identified as the mass of the
black hole. This is also confirmed by evaluating the ADM mass.

We shall prove that $\mu$ is in fact the ADM mass.
According to Mann\cite{M}, the conserved ADM mass $M$ is constructed from the conserved current
\begin{equation}
J_\mu = T_{\mu\nu} \epsilon^{\nu\alpha}\nabla_\alpha K(\phi)
\end{equation}
where $T_{\mu\nu}$ is the stress-energy tensor and $K$ takes the following 
form in our case.  
\begin{equation}
K = K_0 \int^\phi d\phi e^{-4 \int^\phi G} 
\label{Killing}
\end{equation}
where the constant $K_0$ is chosen by the condition that the vector 
$ \xi^\mu = \epsilon^{\nu\alpha} \nabla_\alpha K$ is normalized unity 
at spatial infinity. Then the mass function $m$ is introduced as 
 $J_\mu = \epsilon^{\, \nu}_\mu \nabla_\nu m$ which is written in our present case with $F = \phi$ as follows.
\begin{equation}
m ={K_0\over 2}
\left[ 4\lambda^2 \int^\phi d\phi U e^{-4 \int^\phi G} 
- (\nabla \phi)^2  e^{-4 \int^\phi G} \right]
\label{massfn}
\end{equation}
The ADM mass is evaluated as the limit of $m$ at the spatial infinity. 
In our static situation for $x^+ > z^+$, the mass function is easily 
evaluated as 
\begin{eqnarray}
m &=& {K_0\over 2}
\left[ \lambda^2 \left({U\over G}\right)^{1/2} 
- e^{-2\rho} {\dot \phi}^2 \left({G\over U}\right)^{1/2} \right] \nonumber \\
&=& {K_0\over 2}{\lambda\over g}{\dot \phi} [ e^{-2\rho} - g^2 ] \nonumber \\
&=& K_0 \lambda \mu
\end{eqnarray}
where we have used the relation
\begin{equation}
U = G \exp \left[8\int^\phi G d\phi \right]
\end{equation}
which is obtained from the vacuum existence condition (\ref{11}), and 
the relations ({\ref{24}), (\ref{26}) ,(\ref{27}) and (\ref{37}).  The normalization constant $K_0$ is fixed as follows. 
Using the above relation, (\ref{Killing}) takes the form 
\begin{equation}
K = K_0 \int^\phi d\phi \left({G\over U}\right)^{1/2},
\end{equation}
This gives us the timelike Killing vector $\xi$ whose 
component at the spatial infinity is 
\begin{equation}
{\dot K}(\infty)
= K_0 \left({{\bar G}\over U}\right)^{1/2}|_{r=\infty} 
= K_0 {\lambda\over g}
\end{equation}
Since the metric takes the form $e^{2C} = g^{-2}$ at the spatial infinity, 
the normalization constant is chosen as $ K_0 = 1/\lambda$ 
which gives us the required result for the mass.

\section{Black Holes with Larger Mass}

\ \\

As seen in the section 4,
 giving a black-hole configuration and specifying $g$ and 
 the mass $\mu$  is equivalent to fixing the action of 
the dilaton gravity possessing the minkowski and the black-hole
 configurations as solutions of the equations of motion.

In this section we prove that the theory has static black-hole
 solutions with larger masses than $\mu$.

Taking the dilaton definition as $F=\phi$,
 the following relations come from equations (\ref{24}), 
(\ref{25}) and (\ref{26}).
\begin{eqnarray}
&&
\phi=\phi_o +\int^r_{-\infty} \frac{2g\mu }{ e^{-2\rho }-g^2} dr
\label{50}\\
&&
r=r(\phi -\phi_o , \mu , g)
\\
&&
G=G[ \phi -\phi_o ,\mu ,g,\rho(r)]
\\
&&
U=U[ \phi -\phi_o ,\mu ,g, \rho(r)]
\end{eqnarray}
Let us consider 
another static configuration $(\rho'(r') ,\phi'(r'))$
 as a solution of the same theory; $(F=\phi, G(\phi), U(\phi))$.
 From the same argument in the section 4 
 the following relations must hold.
\begin{eqnarray}
&&
\frac{d\phi'}{dr'}=\frac{2g b M}
{ g^2 e^{-2\rho' }-b^2} ,
\\
&&
r'=r' (\phi' -{\phi'}_o ,M, b ),
\\
&&
G[ \phi' - {\phi'}_o ,M,b,\rho'(r')]
=\frac{b}{4g M}\frac{d\rho'}{dr'},
\\
&&
U[ \phi' - {\phi'}_o ,M,b,\rho'(r')]
=g^3 b \frac{M}{\lambda^2}\left[
\frac{e^{-2\rho'} }{g^2 e^{-2\rho'} -b^2 }
\right]^2
\frac{d\rho'}{dr'},
\end{eqnarray}
where $b$ and $M$ are some constants.

In order to ensure that the 
 configurations $\rho(r)$ and $\rho'(r')$ are the solutions
of the same system, 
 the following relations are needed.
\begin{eqnarray}
&&
G[ \phi' - {\phi'}_o ,M,b,\rho'(r')]
=G[ \phi -\phi_o ,\mu ,g,\rho(r)],
\\
&&
U[ \phi' -{\phi'}_o ,M ,b, \rho'(r')]
=U[ \phi -\phi_o ,\mu, g, \rho(r)].
\end{eqnarray}

These yield two equations:
\begin{eqnarray}
&&
\frac{dr'}{dr}
=\frac{ b \mu }{g^2 M }\frac{d\rho'}{d\rho},
\label{51}\\
&&
e^{2\rho'}
=\frac{g^2}{b^2}
\left[
1-\frac{M}{\mu} (1-g^2 e^{2\rho} )
\right].
\label{52}
\end{eqnarray}

By substituting (\ref{52}) into (\ref{51}) and integrating it 
 we acquire the transformation relation between $r$ and $r'$.
\begin{eqnarray}
r' -{r'}_o
=b\int
\frac{e^{2\rho}}
{1-\frac{M}{\mu}(1-g^2 e^{2\rho} )}dr.
\end{eqnarray}

Now let us concentrate on the parameter space with
$$
M>\mu.
$$
Then there exists a point $r=r_H$ where 
\begin{eqnarray}
e^{2\rho(r_H) } =\frac{1}{g^2} 
\left[ 1-\frac{\mu }{ M}\right].
\end{eqnarray}

When $r$ approaches to $r_H$, the value of $r$ diverges
 to $-\infty$. If we take 
$$
b=g^2 \frac{ M}{\mu\lambda }
\frac{d\rho}{dr}(r_H),
$$
the asymptotic behavior is given as follows.
\begin{eqnarray}
\rho'(r'\sim-\infty)
=\lambda r' +{\rho'}_o +{\rho'}_1 e^{2\lambda r'}
+\cdots .
\end{eqnarray}
This is precisely a black-hole configuration. 
Repeating the junction argument in the section 5 
for this black hole, $M$ can be exactly proven  as 
the mass of the hole.

Note that the regularity of the dilaton field everywhere outside the horizon 
and the permission to have arbitrary heavy mass of the black hole 
are guaranteed by setting  
$$
g=e^{-\rho(\infty )}.
$$
which is derived by the equations (\ref{50}) and (\ref{52}).

Then the functions specifying the action, $F$, $G$ and $U$
are  given by a parametric representation as follows.
\begin{eqnarray}
&&
F(\phi)=\phi\label{500}\\
&&
G(\phi)=\frac{e^{-\rho(\infty)} }{4\mu} \dot{\rho} (r) ,
\label{53}\\
&&
U(\phi)=
\frac{\mu}{\lambda^2}
\frac{e^{-\rho(\infty)}\dot{\rho}(r)}
{(1-e^{2\rho(r) -2\rho(\infty)} )^2} .
\label{54}\\
&&
\phi=\phi_o+2\mu \int_{-\infty}^r 
\frac{e^{-\rho(\infty)} }
{e^{-2\rho(r)} -e^{-2\rho(\infty)} }
dr ,
\label{55}
\end{eqnarray}

Finally we obtain a parametric representation of 
the black-hole solution: 
\begin{eqnarray}
&&
\phi' (r')=\phi(r)
=\phi_o +\int^r_{-\infty} \frac{2\mu e^{-\rho(\infty)}}
{e^{-2\rho(r) }-e^{-2\rho(\infty)} } dr ,
\label{56}\\
&&
e^{2\rho' (r')}
=\frac{\lambda^2 \mu^2 e^{2\rho(\infty)}}{ M^2 \dot{\rho} (r_H )^2 } 
\left[1-\frac{M}{\mu}(1-e^{2\rho(r)-2\rho(\infty)} ) \right],
\label{57}\\
&&
r' -{r'}_o
=
\frac{M}{\mu}\frac{\dot{\rho}(r_H )}{\lambda}
\int \frac{e^{2\rho(r)-2\rho(\infty)} }{1-\frac{M}{\mu}
( 1 - e^{2\rho(r) -2\rho(\infty)} ) }dr ,
\label{58}
\end{eqnarray}
where 
$$
e^{2\rho(r_H )-2\rho(\infty)} =1-\frac{\mu}{M} .
$$

This is our main result. Once the input black-hole 
configuration $\rho(r)$ and its mass $\mu$ are given arbitrary,
 the theory is fixed by the equations (\ref{500}), (\ref{53}) 
and (\ref{54}). Then the equations (\ref{56})$\sim$
(\ref{58}) generate  black holes with $M>\mu$.
\ \\

\section{CGHS Black Holes}

\ \\

In this section we check the formula (\ref{53})$\sim$(\ref{58})
for the well-known CGHS black hole\cite{CGHS}.

Let first give the CGHS black hole with mass $\mu =\lambda$.
\begin{eqnarray}
e^{2\rho}
=\frac{1}{1+e^{-2\lambda r}}.
\end{eqnarray}

Then the horizon for the black hole with mass $M$ is evaluated
as follows.
\begin{eqnarray}
&&
r_H 
=-\frac{1}{2\lambda}\ln
\left(
\frac{ \lambda}{M-\lambda} 
\right)
\nonumber\\
&&
\dot{\rho}(r_H )
=\frac{\lambda^2}{ M}
\nonumber
\end{eqnarray}

The equation (\ref{55}) with 
$$
\phi_o =1
$$
 yields
\begin{eqnarray}
&&
\phi =1 +e^{2\lambda r} ,
\\
&&
r=\frac{1}{2\lambda}\ln [\phi -1].
\end{eqnarray}

The equation (\ref{53}) and (\ref{54}) give 
\begin{eqnarray}
&&
G=\frac{1}{4}\frac{1}{1+e^{2\lambda r}}=\frac{1}{4\phi}
\nonumber\\
&&
U=1+e^{2\lambda r} =\phi.
\nonumber
\end{eqnarray}
Thus equation (\ref{53})$\sim$ (\ref{55}) reproduce the
precise form of the CGHS action.

From equations (\ref{55})$\sim$ (\ref{57}) with
$$
{r'}_o =0,
$$
the following relations come out.
\begin{eqnarray}
&&
r' 
=
\frac{1}{2\lambda }
\ln \left[e^{2\lambda r} +1-\frac{M}{\lambda} \right]
,
\nonumber\\
&&
e^{2\rho'}
=\frac{1}{1+\frac{M}{\lambda}e^{-2\lambda r'} },
\nonumber\\
&&
\phi' = e^{2\lambda r'}  + \frac{M}{\lambda}  .
\nonumber
\end{eqnarray}
Thus the CGHS black-hole  solutions with mass $M$ 
is exactly reproduced.

\section{Modified CGHS Black Holes}

\ \\

In this section we discuss a more non-trivial example.
Let us set up the metric like 
$$
e^{2\rho}
=\frac{1+\epsilon e^{2\lambda r} }
{1+e^{-2\lambda r} +\epsilon e^{2\lambda r} } .
$$

Also let its mass 
$$
\mu =\lambda 
$$
 and take the parameters as follows.
$$
\phi_o =1,
$$
$$
{r'}_o =0 .
$$

From equation (\ref{55}) the dilaton field is solved as
\begin{eqnarray}
&&
\phi(r) =1+e^{2\lambda r} +\frac{\epsilon}{2}e^{4\lambda r},
\label{61}
\\
&&
r=\frac{1}{2\lambda}
\ln
\left[
\frac{1}{\epsilon}
(
\sqrt{1+2\epsilon (\phi -1) }-1
)
\right].
\label{62}
\end{eqnarray}

Substituting equation(\ref{62}) into equations (\ref{53})
 and (\ref{54}) yields
\begin{eqnarray}
&&
F(\phi)=\phi ,
\label{93}
\\
&&
G(\phi)=\frac{1}{4}
\frac{
\left[
\sqrt{1+2\epsilon (\phi -1)}-\frac{1}{2}
\right]
\left[
1+\epsilon (2\phi -1)
+\sqrt{1+2\epsilon (\phi -1)}
\right]
}
{
\left[
\phi +\frac{\epsilon}{2}(2\phi -1)^2
\right]
\sqrt{1+2\epsilon (\phi -1)}
} ,
\label{94}\\
&&
U(\phi)=\frac{4}{\epsilon^2}G(\phi)
\left[
1+\epsilon (2\phi -1)
-\sqrt{1+2\epsilon (\phi -1)}
\right]^2 .
\label{95}
\end{eqnarray}
These determine the action of the theory.

Next consider the black holes with mass $M>\mu$.
The horizon is given as follows.
\begin{eqnarray}
r_H =
\frac{1}{2\lambda}
\ln \left[\frac{1}{2\epsilon}
\left[
\sqrt{1+4\epsilon \left(\frac{M}{\lambda} -1 \right)}-1
\right] \right],
\end{eqnarray}
$$
\dot{\rho}(r_H  )
=
\frac{\lambda^3}
{2\epsilon M(M-\lambda)}
\sqrt{1+4\epsilon \left( \frac{M}{\lambda} -1 \right)}
\left[
\sqrt{1+4\epsilon \left( \frac{M}{\lambda} -1 \right)}
-1
\right] .
$$

Then the final result 
is expressed as follows.
\begin{eqnarray}
&&
\phi =1+s +\frac{\epsilon}{2}s^2 ,
\nonumber\\
&&
e^{2\rho}
=
\frac{1+2\epsilon\left(\frac{M}{\lambda}-1 \right)+
\sqrt{1+4\epsilon \left(\frac{M}{\lambda}-1 \right)}}
{2\left(1+4\epsilon \left(\frac{M}{\lambda}-1  \right) \right)}
\frac{s+\epsilon s^2+1 -\frac{M}{\lambda} }
{1+s+\epsilon s^2} ,
\nonumber\\
&&
e^{2\lambda r'}
=
\left[s +\frac{1}{2\epsilon}
\left(1-\sqrt{1+4\epsilon \left(\frac{M}{\lambda} -1 \right)}
\right)
\right]^{1-\beta}
\left[s+\frac{1}{2\epsilon}
\left(1+\sqrt{1+4\epsilon \left(\frac{M}{\lambda} -1 \right)}
\right)
\right]^\beta ,
\nonumber\\
&&
\beta =\frac{1}{2}
\left[
1-\frac{1}{\sqrt{1+4\epsilon\left( \frac{M}{\lambda}-1 \right)}}
\right] .\nonumber
\end{eqnarray}
where $s=e^{2\lambda r}$.

This is a parametric representation of the black-hole solution
with mass $M$ in the theory with equations 
(\ref{93})$\sim$(\ref{95}). 
The parameter $s$ runs in the following region.
\begin{eqnarray}
\frac{1}{2\epsilon}
\left[
\sqrt{
1+4\epsilon \left(\frac{M}{\lambda} -1 \right)
}-1
\right]
&\leq& s \leq \infty .
\nonumber\\
(-\infty &\leq& r \leq \infty .)
\nonumber
\end{eqnarray}

\section{Summary}

We have investigated the conditions for two-dimensional dilaton gravity 
to have a vacuum solution as well as a black hole solution. 
The general expression for a black hole solutions is obtained by giving an arbitrary function with appropriate boundary conditions.
Then we derive the solution describing  dynamical formation of a black hole 
by a shock-wave type material source.  Furthermore we obtained a parametric 
representation of general solution of black holes with arbitrary mass. 
Thus our work should be regarded as the general formalism to 
derive general solutions describing not only a static black hole but also dynamical formation of black hole. 
It is quite interesting to see the quantum effect in our model. 
This remains as a future problem.

\section*{Acknowledgments}
We would like to thank M. Morikawa for fruitful discussions. 
This work is supported in part by Japanese Grant-in-Aid for Science 
Research fund of Ministry of Education, 
Science and Culture No.09640332(T.F).

\end{document}